\documentclass[12pt]{article}

\columnwidth\textwidth

\begin{document}

\begin{titlepage}

\noindent
\hspace*{11cm} BUTP-99/05 \\
\vspace*{1cm}
\begin{center}
{\LARGE Choice of Gauge in Quantum Gravity}    

\vspace{1cm}

P. H\'{a}j\'{\i}\v{c}ek \\
Institute for Theoretical Physics \\
University of Bern \\
Sidlerstrasse 5, CH-3012 Bern, Switzerland \\
\vspace*{1cm}

March 1999 \\ \vspace*{.5cm}

\nopagebreak[4]

\begin{abstract}
  This paper is an extended version of the talk given at 19th Texas Symposium
  of Relativistic Astrophysics and Cosmology, Paris, 1998. It reviews of some
  recent work; mathematical details are skipped. It is well-known that a
  choice of gauge in generally covariant models has a twofold pupose: not only
  to render the dynamics unique, but also to define the spacetime points. A
  geometric way of choosing gauge that is not based on coordinate
  conditions---the so-called covariant gauge fixing---is described. After a
  covariant gauge fixing, the dynamics is unique and the background manifold
  points are well-defined, but the description remains invariant with respect
  to all diffeomorphisms of the background manifold.  Transformations between
  different covariant gauge fixings form the well-known Bergmann-Komar group.
  Each covariant gauge fixing determines a so-called Kucha\v{r} decomposition.
  The construction of the quantum theory is based on the Kucha\v{r} form of
  the action and the Dirac method of operator constraints. It is demonstrated
  that the Bergmann-Komar group is too large to be implementable by unitary
  maps in the quantum domain.
\end{abstract}

\end{center}

\end{titlepage}

\section{Introduction}

In this paper, we are considering a broad class of diffeomorphism invariant
models similar to general relativity. Thus, the dynamical equations will
be generally covariant and the spacetime will be dynamical.

In such a situation, one might be tempted to view the system as a dynamics of
some fields and objects on a naked manifold, the manifold consisting of
well-defined and distinguishable points. This point of view is, however,
afflicted with well-known difficulties and paradoxes. One old example of such
difficulties is Einstein's `hole' argument \cite{hole}, whereas a more
recent example is due to Fredenhagen and Haag \cite{F-H}. The way out of the
difficulties was known already to Einstein \cite{hole}: spacetime points can
only be defined and distinguished by values of physical fields or positions of
physical objects.

This principle is not spectacular if one does not leave the realm of a fixed
classical solution, but it is rather awkward from the point of view of the
whole dynamics of the model, especially if one is interested in its
quantization. Various methods can be found in the literature that help to
circumvent the problem. One is a WKB expansion around a (classical) solution;
this enables one to define the spacetime points by means of the classical
metric and fields of the solution similarly as it is done in Minkowski
spacetime. Of course, this method will work only if the WKB approximation is
applicable.  Another method is to add some material that breaks the
diffeomorphism invariance; this way has been quite systematically explored by
Kucha\v{r} \cite{kuch}. Finally, one can, so to speak, fasten the coordinates
to particular bumps of the fields in the model, that is, one chooses a gauge.
This last method seems to be, unlike the first one, generally applicable, and
it does not, like the second one, violate the diffeomorphism (gauge)
invariance, provided one can prove that the measurable results are independent
of the gauge choice.

In the present paper, we shall concentrate on the last method and study the
question of how much it can be used in the quantum theory of the generally
covariant models. Our results will suggest that quantum theories constructed
on the basis of different gauges are not unitarily equivalent. The reason is
that their gauge group is huge; it is not just the diffeomorphism group of one
manifold, but a cartesian product of diffeomorphism groups, one group for the
spacetime of each solution; this is the well-known Bergmann-Komar group
\cite{B-K}.

\section{Covariant Gauge Fixing}
\label{sec:CGF}
In this section, we explain the origin of the Bergmann-Komar group and sketch
the idea of our definition of the covariant gauge fixing.

It is clear that each fixed classical spacetime solution defines a nice
spacetime; problems can only arise if one considers more than one
solution. To see this, we study a simplified example: the family of
Schwarzschild spacetimes. Let us first consider the metric in the
Eddington-Finkelstein coordinates:
\begin{equation}
  ds^2 = -\left(1-{2M\over R}\right)dW^2 + 2dWdR + R^2d\Omega^2,
\label{EF}
\end{equation}
where $d\Omega^2$ is the metric of the unit sphere in the spherical
coordinates $\vartheta\in (0,\pi)$ and $\varphi\in (0,2\pi)$;
$W\in(-\infty,\infty)$ and $R\in(0,\infty)$. $M$ is the Schwarzschild mass:
For each value of $M\in(0,\infty)$, there is one spacetime manifold and a
metric on it. Eq.\ (\ref{EF}) can also be interpreted as a family of metric
fields on a fixed background manifold ${\mathcal M}_{EF}$; this is simply the
manifold ${\mathbf R}^2\times S^2$ covered by the coordinates $W$, $R$,
$\vartheta$ and $\varphi$ with the above ranges.

Another possibility is to choose the Kruskal coordinates. Then, the metric has
the form
\begin{equation}
  ds^2 = -{16M^2\over\kappa(-UV)}e^{-\kappa(-UV)}dUdV
  +4M^2(\kappa(-UV))^2d\Omega^2, 
\label{K}
\end{equation}
where $\kappa : (-1,\infty) \mapsto (0,\infty)$ is the well-known Kruskal
function defined by its inverse, $\kappa^{-1}(x) := (x-1)e^x$ for all $x\in
(0,\infty)$; $U\in (-\infty,\infty)$ and $V\in (0,\infty)$. Again, we can
interpret Eq.\ (\ref{K}) as a family of metric fields on a background manifold
${\mathcal M}_K$; it is ${\mathbf R}^2\times S^2$ with coordinates $U$,
$V$, $\vartheta$ and $\varphi$ in the above ranges.

The crucial observation is that the transformation between the
Eddington-Finkelstein and Kruskal coordinates,
\begin{eqnarray*}
  U & = & \left({R\over 2M} - 1\right)e^{R/2M}e^{-W/4M}, \\
  V & = & e^{W/4M},
\end{eqnarray*}
is {\em not} a map from ${\mathcal M}_{EF}$ to ${\mathcal M}_K$ because it
depends on $M$. The conclusion from this observation is that the background
manifold is not defined by any unique and natural way, but by some non-trivial
method of identifying all solution manifolds. In our case, we have introduced
a particular geometric coordinate system in each solution manifold, and then
identified those points of different solution manifolds that have the same
values of these coordinates. The choice of particular geometric coordinates in
all solutions is usually a consequence of a gauge choice (or coordinate
condition).  Hence, it is the gauge fixing that defines points of a background
manifold in general relativity. Moreover, a transformation between two
different gauge fixings is not a transformation of coordinates on the
background manifold, but a whole {\em family} of coordinate transformations,
one for each solution.  That is the origin of Bergmann-Komar group.

Turning to a general case, we need some additional notions. Our description of
a model is based on the canonical formalism. Each model possesses an extended
phase space $\Gamma'$ that contains a constraint surface $\Gamma$. Each point
of $\Gamma$ represents a possible initial datum for the dynamical equations of
the model. Points at $\Gamma$ that represent initial data of one and the same
solution form a surface in $\Gamma$ that we call c-orbit. The c-orbits are in
one-to-one correspondence with classical solutions. The Bergmann-Komar group
acts only at $\Gamma$. It is generated by vector fields at $\Gamma$ that are
tangential to c-orbits.

The discussion above of the Schwarzschild family seems to imply that it is not
necessary to choose coordinates in order to define the background manifold;
one can also directly identify the points of different solution spacetimes. An
advantage thereof is that no solution has to be covered by just one chart;
another one is that everything can be done in a way covariant with respect to
the coordinate transformation on the resulting background manifold. A crucial
condition for such a construction to work is the absence of the points at the
constraint surface that correspond to spacetime solutions with {\em any}
symmetry (even a discrete one). We cut these points out; then there is a
one-to-one relation between remaining points and Cauchy surfaces in the
solutions. Suppose that, for each solution, there is a smooth injection
sending this solution into a fixed background manifold. Then, for each point of
the constraint surface, there is a unique embedding of the Cauchy manifold
into the background manifold: these embeddings are made to functions on the
constraint surface. One can then show that the embeddings together with any
complete system of functions that are constant along the c-orbits form a
coordinate system at $\Gamma$. The set of the differentiable injections
described above is called {\em covariant gauge fixing}.

Such a construction has been performed in detail for a system that can be
called ``extended shell model'' \cite{shell}. It consists of a null-dust thin
shell surrounded by its own gravitational field; everything is spherically
symmetric. This is a system of one degree of freedom (the radius of the shell,
say), and one usually reduces the action correspondingly, see, e.g.\ 
\cite{louko}. The word ``extended'' in the name of the model refers to the
spacetime outside. Such extension is necessary in order that a generally
covariant model with a dynamical spacetime results.

An important question about covariant gauge fixing is that of its global
existence. This existence has been shown as yet for the following models: the
minisuperspace cosmological models, 2+1 gravity \cite{moncrief}, the
cylindrical waves \cite{cylinder}, the Schwarzschild family \cite{kul}, a
dilatonic model \cite{KRV}, and the extended shell model \cite{shell}. For
general relativity, the existence may be a problem. However, if a global
covariant gauge fixing does not exist for some model, then the model cannot be
reformulated as a field theory on a background manifold even in its classical
version and one does not need to pass to quantum theory in order to prove this
negative result.

\section{The Kucha\v{r} Decomposition}
\label{sec:KD}
The Kucha\v{r} decomposition will play an important role in our argument. Let
us briefly introduce it using the Hamiltonian formulation of general
relativity.

The Hamiltonian formalism starts with the so-called ADM action:
\[
  S = \int dt\int_\Sigma d^3x\left[\pi^{kl}(x)\dot{q}_{kl}(x) - {\mathcal N}(x)
  {\mathcal H}(x) - {\mathcal N}^k(x){\mathcal H}_k(x)\right],
\]
where $\Sigma$ is a three-dimensional initial value manifold (Cauchy surface),
the pair of fields $(q_{kl}(x),\pi^{kl}(x))$ on $\Sigma$ determines a point of
$\Gamma'$, ${\mathcal N}(x)$ and ${\mathcal N}^k(x)$ are Lagrange multipliers
and ${\mathcal H}(x)$ and ${\mathcal H}^k(x)$ are the constraints; the
constraints are functionals of the fields $q_{kl}(x)$ and $\pi^{kl}(x)$.
$\Gamma$ is determined by the constraint equations ${\mathcal H}(x) = 0$ and
${\mathcal H}^k(x) = 0$.

From the above action, it follows that the dynamics is generated by the
constraints; the Lagrange multipliers can be interpreted as components of the
vector fields determining the direction in which the dynamics proceeds (for
details, see \cite{MTW}). The functionals $F[q_{kl}(x),\pi^{kl}(x)]$ that
satisfy the conditions
\[
  \{F,{\mathcal H}\}|_\Gamma = 0,\quad \{F,{\mathcal H}_k\}|_\Gamma = 0,
\]
are gauge invariants and simultaneously integrals of motion, so they are
constant along the c-orbits; we call them
{\em perennials}.

Kucha\v{r} observed (for some simplified cases) \cite{cylinder,3way} that
there is a canonical transformation from the variables
$(q_{kl}(x),\pi^{kl}(x))$ to new variables
$(X^\mu(x),P_\mu(x),q_\alpha,p^\alpha)$ such that the action becomes
\begin{equation}
  S = \int dt\left[\int_\Sigma d^3x\left(P_\mu(x)\dot{X}^\mu(x) - {\mathcal
  N}^{\prime\mu}(x)P_\mu(x)\right) + \sum_\alpha p_\alpha
  \dot{q}^\alpha\right],
\label{KD}
\end{equation}
where $X^\mu(x)$ is a coordinate description of an embedding of the Cauchy
surface $\Sigma$ into some background manifold $\mathcal M$, $X^\mu$ being
some coordinates on $\mathcal M$, $x^k$ those on $\Sigma$, and
$(q^\alpha,p_\alpha)$ represent some complete system of perennials; the index
$\alpha$ can run through discrete and/or continuous ranges, and the sum in the
action is, therefore, to be understood as a kind of Stieltjes integral.

The new variables can be cleanly split into pure kinematical variables
$X^\mu(x)$ and $P_\mu(x)$ on one hand, and true dynamical variables $q_\alpha$
and $p^\alpha$ on the other---this is what we call {\em Kucha\v{r}
  decomposition}.

Each value of the pair $(q^\alpha,p_\alpha)$ from some range determines a
unique solution; the metric of the solution can be written as a
$(q^\alpha,p_\alpha)$-dependent metric field on the background manifold
$\mathcal M$:
\begin{equation}
  ds^2 = g_{\mu\nu}(q,p;X)dX^\mu dX^\nu.
\label{KRV}
\end{equation}
This metric appeared first in \cite{KRV} and we call it
Kucha\v{r}-Romano-Varadarajan (KRV) metric. Every Kucha\v{r} decomposition
must clearly be associated with some gauge fixing.

\section{Extension Theorem}
\label{sec:ET}
In Sec.\ \ref{sec:CGF}, we have seen that a covariant gauge fixing leads to a
definition of the coordinates $X^\mu(x)$, $q^\alpha$ and $p_\alpha$ on the
constraint surface $\Gamma$. According to the previous section, this is a
Kucha\v{r} decomposition at $\Gamma$. In the present section, we show that
there is a Kucha\v{r} decomposition in $\Gamma'$ for each covariant gauge
fixing.

One would like that the following properties hold:
\begin{enumerate}
\item The functions $X^\mu(x)$, $q^\alpha$ and $p_\alpha$ as constructed by a
  covariant gauge fixing in Sec.\ \ref{sec:CGF} can be extended to a
  neighbourhood $U$ of $\Gamma$ in $\Gamma'$ so that their Poisson brackets
  in $U$ are
  \[ \{X^\mu(x),X^\nu(y)\} = 0,\quad \forall \mu,\nu,x,y, \]
  \[ \{X^\mu(x),q^\alpha\} = \{X^\mu(x),p_\alpha\} = 0,\quad \forall
  \mu,x,\alpha, \]
  \[ \{q^\alpha,p_\beta\} = \delta^\alpha_\beta,\quad \{q^\alpha,q^\beta\} =
  \{p_\alpha,p_\beta\} = 0,\quad \forall \alpha, \beta. \]
\item There are functions $P_\mu(x)$ in $U$ such that the equations $P_\mu(x)
  = 0$ define $\Gamma$, and such that
\item the Poisson brackets of $P$'s with the other functions in $U$ are:
  \[ \{P_\mu(x),P_\nu(y)\} = 0,\quad \forall \mu,\nu,x,y, \] 
  \[ \{P_\mu(x),q^\alpha\} = \{P_\mu(x),p_\alpha\} = 0,\quad \forall
  \mu,x,\alpha, \]
  and 
  \[ \{X^\mu(x),P_\nu(y)\} = \delta^\mu_\nu\delta(x,y),\quad \forall
  \mu,\nu,x,y. \] 
\end{enumerate} 
The proof of these properties is based on the Darboux-Weinstein theorem (see,
e.g.\ \cite{zielon}). A formal proof is easy; ``formal'' means that the
subtleties of infinite-dimensional spaces are neglected. A full (rigorous)
proof is given for the extended shell model \cite{shell}; it uses the ideas
about weak symplectic forms and the associated weak metrics as described in
\cite{marsden}. Extension of the proof to other models seems to be
straightforward, if some quite plausible assumptions about the submanifold
structure of the constraint set $\Gamma$ and of the quotient manifold
structure of the set $\Gamma/$c-orbits (true degrees of freedom) are
satisfied.

The theorem is a pure existence theorem, even if its proof provides a method
of how the extensions of $X^\mu(x)$, $p_\alpha$ and $q^\alpha$ out of the
constraint surface can be constructed, at least in principle; the method is
practically viable only in very simple cases. Still, the result is useful
because no explicit knowledge of the extension is, in fact, needed for the
construction of the quantum theory. An explicit calculation of the extension
by any known method is very difficult. Kucha\v{r} and his collaborators
managed to find such extensions only in the cases of cylindrical gravitational
waves \cite{cylinder}, spherically symmetric vacuum gravitational field
\cite{kul}, and a dilatonic model of gravitational collapse \cite{KRV}.

The extensions guaranteed by the theorem enable us to extend the covariant
gauge fixing and the Bergmann-Komar group out of the constraint surface. This,
however, is not really necessary.  Further, it is even not unique, because the
extensions are not (this follows from the proof). An important property of the
Bergmann-Komar transformation is that it does not change the values of the
perennials $q^\alpha$ and $p_\alpha$ at the constraint surface $\Gamma$; as it
is well-known, the Poisson algebra of perennials is determined by their
restrictions to $\Gamma$ \cite{marsden}, so this algebra is also gauge
invariant.

\section{Quantum Theory}
The construction of the quantum theory based on the action in the
Kucha\v{r} form (\ref{KD}) and the Dirac method of operator constraints is
rather straightforward and we shall only briefly sketch it.

The states are described by wave functions $\Psi(X,q)$ of the embeddings
$X^\mu(x)$ and the perennials $q^\alpha$. The functional Schr\"{o}dinger
equation (see, e.g.\ \cite{canada}) then reads:
\[
  \hat{P}_\mu(x)\Psi = {1\over i}{\delta \Psi\over \delta X^\mu(x)} = 0.
\]
Hence, physical state $\Psi$ is simply independent of the embeddings. 

The resulting quantum theory possesses what we can call a {\em gauge invariant
  core}. This consists of (i) the states described by wave functions
$\Psi(q)$, (ii) the scalar product,
\[
  (\Psi,\Phi) := \int d\mu(q)\,\Psi^*(q)\Phi(q),
\]
where $d\mu(q)$ is some measure on the space with coordinates $q^\alpha$, and
(iii) the observables $\hat{q}^\alpha$ and $\hat{p}_\alpha$ defined by
\[
  \hat{q}^\alpha\Psi(q) := q^\alpha\Psi(q),\quad \hat{p}_\alpha\Psi(q) :=
  {1\over i}{\partial \Psi \over \partial q^\alpha}.
\]
This structure is completely independent of any choice of gauge (or covariant
gauge fixing), and it is manifestly invariant with respect to the
Bergmann-Komar group.

In principle, there is also additional information about the points and
geometry of the spacetime. At least in the classical theory, the spacetime
points could be defined by a covariant gauge fixing, and the KRV metric
(\ref{KRV}) determined the corresponding geometric properties. In the
classical theory, this description of the properties is gauge dependent, but
the results for really observable properties are gauge independent. This part
of the classical theory would correspond to the method of gauge fixing known
from other gauge theories. For example the values of coordinates after a gauge
has been fixed are, in principle, measurable quantities. In our quantum
theory, the embeddings $X^\mu(x)$ are trivial operators: they commute with
every other variable and with each other (the conjugate quantities $P_\mu(x)$
have been excluded from the quantum theory). This corresponds to the situation
that is usual in the quantum field theory: the spacetime coordinates are just
parameters.  We asume that the KRV metric can be transferred to the quantum
theory by some suitable factor ordering.  The resulting {\em operator of
  geometry} $\hat{g}_{\mu\nu}(\hat{q},X)$ depends on the four parameters
$X^0$, $X^1$, $X^2$ and $X^3$ and represents the ``quantum geometry'' at the
point $(X^0, X^1, X^2, X^3)$ of the background manifold $\mathcal M$. This
formalism can be made manifestly invariant with respect to the group of
diffeomorphisms of $\mathcal M$, or with respect to any coordinate
transformation $X^{\prime\mu} = X^{\prime\mu}(X)$ on $\mathcal M$, but not
with respect to the full Bergmann-Komar group. In fact, we show that general
Bergmann-Komar transformations cannot be unitarily implemented for our quantum
theories.

Let us choose two different covariant gauge fixings; these lead to two
different descriptions $X_1^\mu(x)$ and $X_2^\mu(x)$ of embeddings, and to two
Kucha\v{r} decompositions. Using these decomposition, we can construct the
corresponding quantum theories $QT_1$ and $QT_2$ by the method given
above. Within the theory $QT_1$, $X_1^\mu$ are $c$-numbers, within $QT_2$,
$X_2^\mu$ are. However, $X_2$ in $QT_1$ is given by the
Bergmann-Komar transform:
\[
  X_2^\mu = X_2^\mu(X_1^\mu,q^\alpha,p_\alpha).
\]
Hence, within $QT_1$, $X_2$ is a genuine operator; formally, 
\[
  \hat{X}_2^\mu := X_2^\mu(X_1^\mu,\hat{q}^\alpha,\hat{p}_\alpha)
\]
(we suppose that the corresponding factor ordering problem can be solved
satisfactorily). Thus, $X_2$ is a $q$-number in $QT_1$ and a $c$-number in
$QT_2$.  If $QT_1$ and $QT_2$ are unitarily equivalent, then the unitary map
between the corresponding Hilbert spaces must send the physical quantities
with the same meaning into each other. For example, $\hat{X}^\mu_2$ of $QT_2$
is to be sent into $\hat{X}^\mu_2$ of $QT_1$. However, no unitary map can send
$c$-numbers in $q$-numbers.

\section{Conclusion}
We have studied the symplectic structure of diffeomorphically invariant
models, especially the aspects associated with gauge (that is, coordinate)
choice. There were two main results.

The first one is the existence of the Kucha\v{r} decomposition for each
covariant gauge fixing. The corresponding quantum theory also decomposes,
namely into a gauge-invariant core in the form of a representation of an
algebra of observables on one hand, and into the additional information about
geometry containing the definition of background manifold and its points, and
the quantum geometry at these points, on the other.

The second main result is that the information in the second part of the
quantum theory depends of the gauge. This suggests that, if we insist on gauge
invariance in quantum gravity, there may be no spacetime points, and that
gauge fixing methods may be more precarious than, say, in quantum Yang-Mills
field theories.

\section*{Acknowledgments}
I would like to thank J.~Kijowski for bringing the Darboux-Weinstein theorem
to my attention, and to J. Bi\v{c}\'{a}k, H. Friedrich and J. Whelan for
helpful discussions.

\end{document}